\input{aipcheck}

\documentclass[
    ,final            
    ,numberedheadings 
  ]
  {aipproc}

\usepackage{bm}
\layoutstyle{6x9}

\begin{document}

\title{Forecasting Seismic
Signatures of Stellar Magnetic Activity}



\author{W. A. Dziembowski}{
  address={Warsaw University Observatory, Aleje Ujazdowskie 4, 00-478
  and Copernicus Astronomical Center, ul. Bartycka 18, 00-716 Warszawa, Poland}
}

\begin{abstract}
For the Sun, a tight correlation between various activity measures
and oscillation frequencies is well documented. For
other stars, we have abundant data on magnetic activity and its
changes but not yet on its seismic signature. A prediction of
the activity induced frequency changes in stars based
on scaling the solar relations is presented. This seismic
signature of the activity should be measurable in the data expected
within few years.

\end{abstract}

\maketitle


\section{Introduction}

Only a few years after the discovery of low-degree modes of the solar
five minute oscillations, Woodard and Noyes (1985) showed using
data from the SMM mission taken during the declining phase of
cycle 21 that the oscillation frequencies decrease with decreasing
Sun's activity. The correlation between various activity measures
and the frequencies was definitely confirmed with the data taken
during the next activity cycle (Libbrecht \& Woodard 1990;
Bachmann \& Brown 1993; Regulo et al. 1994; Elsworth et al.
1994). High precision frequency measurements are now available for
a number of solar-like stars (e.g. Bedding \& Kjeldsen 2007).
Although, the expected seismic signature of the magnetic activity
is very weak indeed, its discovery seems to be only a matter of
time.

The relative frequency changes in the eleven-year cycle are at
most of the order of $10^{-4}$, which is one order of magnitude less
than the changes in the irradiance and three orders less than in
the chromospheric emission measures. We should expect a similar
situation in distant stars. For many
solar-like objects, cyclic variations variations of the
chromospheric emission are well established (e.g. Baliunas 1998).
Nonetheless, an effort to measure the much weaker seismic signature
is of interest because it reflects the changes taking place in
deeper layers.

In expectation of data from measurements,
Metcalf et al. (2007, hereinafter M07)
developed a method for predicting frequency shifts based on
scaling the measured p-mode frequency variations and changes
of the chromospheric activity indices in the Sun.
Specific forecast was made for the radial modes in the
subgiant $\beta$ Hyi. In this paper, I outline the principle of
the forecasting method adopted in that work and expand on the part
concerning nonradial modes.
\section{Activity related changes in oscillation frequencies}
The starting point in the forecast presented in M07
was the variational expression for frequency shifts
\begin{equation}
\Delta\nu_j={\int d^3{\boldsymbol x}{\cal K}_j{\cal S}\over
2I_j\nu_j},\label{Dn1}
\end{equation}
where $j\equiv(n,\ell,m)$ identifies the mode, ${\cal K}_j$ is the
kernel, which depends on the unperturbed stellar model parameters and
eigenvectors of its free oscillations, ${\boldsymbol\xi}({\boldsymbol x})$.
The quantity
\begin{equation}
I_j=\int d^3{\boldsymbol x}\rho|{\boldsymbol\xi}|^3
   =R^5\bar\rho\tilde I\label{I}
\end{equation}
denotes the mode inertia. The second equality defines the dimensionless
mode inertia, $\tilde I$. The source, ${\cal S}$, must include the
direct feed-back from the magnetic field perturbed by an oscillation
mode but also the indirect field effects caused by the perturbation of the
equilibrium density, temperature, and velocity of convection. All
these effects may be important for frequency shifts at a level of
$0.1 \mu$Hz.
Disentangling these effects in the solar data is
impossible and calculation of their relative contributions to the
source is still very uncertain. The general form of the kernels
for various effects (see e.g. Goldreich et al. 1991, Dziembowski
\& Goode 2004) is complicated but for all of them the leading
terms are proportional to $|{\mbox{\rm div}}{\boldsymbol\xi}_j|^2$.

In their semi-empirical determination of ${\cal S}({\boldsymbol x})$ for the
Sun, M07 adopted the common kernel
\begin{equation}
{\cal K}_j({\boldsymbol x})=|{\mbox{\rm div}}{\boldsymbol\xi}_j|^2
                           =q_j(D)|Y_\ell^m|^2\label{K1},
\end{equation}
where $D$ denotes the depth below the photosphere, absorbing the
model dependent parameters such as gas density or adiabatic
exponent in the unknown source ${\cal S}$. Only the part of ${\cal
S}$ that is symmetric about the equator affects mode frequencies.
Therefore, without loss of generality, they could use the
following expression
\begin{equation}
{\cal S}({\boldsymbol x})=\sum_{k=0}{\cal S}_k(D)P_{2k}(\mu), \label{S1}
\end{equation}
where $\mu=\cos\theta$. It follows from equations\,\eqref{Dn1}, \eqref{K1},
and \eqref{S1} that only the terms up to $k=\ell$ contribute to the
shifts of modes with degree $\ell$.
\section{Frequency changes in the Sun}
Most of what we know about the changes in the solar oscillation
frequencies comes from disc-resolved measurements. The data from such
measurements are now commonly represented in the form
\begin{equation}
\nu_{nlm}=\sum_{i=0}a_{i,n\ell}{\cal P}^\ell_i(m),\label{a}
\end{equation}
where ${\cal P}^\ell_i(m)$ are orthogonal polynomials of order $i$
(see e.g. Schou et al. 1994). The  mean frequencies,
$a_0$, yield information on the radial structure of the Sun.
The higher even-order $a_k$-coefficients yield separately
information on the Sun's asphericity described by the individual
Legendre polynomial $P_{2k}(\mu)$ . Thus,
from measured shifts, $\Delta a_{2i}$ ($i=0, 1,...$), as functions of
the mean frequency, we get our constraints on
${\cal S}_{2i}(D)$ in the Sun.

Using equations\,\eqref{K1} and \eqref{S1} in \eqref{Dn1}, we get
\begin{equation}
\Delta\nu_{nlm}={q_{nl}\sum_{k=0}^\ell{\cal S}_k\kappa_{k,lm} \over
2I_{nl}\nu_{nl}}\label{Dn2}
\end{equation}
where, $$\kappa_{k,lm}\equiv\int d\mu d\phi|Y_\ell^m|^2P_{2k}(\mu)={\cal
P}^\ell_{2k}(m)Z^\ell_k$$ and $$Z^\ell_k=(-1)^k{(2k-1)!!\over
k!}{(2\ell+1)!!\over(2\ell+2k+1)!!} {(\ell-1)!\over(\ell-k)!}.$$
This equation with ${\cal S}_k$ derived from data extending from $\ell=0$
to over $\ell=100$, allows us to determine the shifts of the low-$\ell$
modes much more accurately than if they may have been measured directly.

Data on the frequency shifts show that the source is localized
near the Sun's photosphere. Thus, the following simple form of the
depth dependence of the coefficients in \eqref{S1} was assumed
in M07:
\begin{equation}
{\cal S}_k(D)=1.5\times10^{-11}A_k\delta(D-D_{c,k}) \mu{\rm Hz}^2,\label{S2}
\end{equation}
where $A_k$ and $D_{c,k}$ were determined by fitting equation (7) to the
data of the $a_{2k}$ coefficient shifts,
using the relation
\begin{equation}
\Delta a_{2k,n\ell}=A_kZ_{k,\ell}Q_{n\ell}(D_{c,k})\label{Da},
\end{equation}
where we denoted
\begin{equation}
Q_{n\ell}=1.5\times10^{-11}{q_{n\ell}\over\nu_{n\ell}\tilde I_{n\ell}}\label{Q}.
\end{equation}
The relation \eqref{Da} follows from equations\,\eqref{a}, \eqref{Dn2},
\eqref{S2},
and \eqref{I}. The numerical factor is arbitrary. M07 determined only
the value of $A_0$, using data for p-modes with $n>1$
and $\nu>2.5$. They showed that the mean frequency
shifts, $\Delta a_0$, for over 1500 p-modes are well described with
our two-parameter fit and that $D_{c,0}\approx0.3\mbox{Mm}$. There was
a small systematic departure at low frequencies, indicating that
the source extends beneath 0.3 Mm, but it was regarded to be unimportant.
For other $a_{2k}$ of low $k$, the fit is also good at similar
values of $D_{c,k}$. Thus, data on shifts for each of the low order
even $a$-coefficients may be reduced to the single parameters
$A_k$. How these parameters correlate with the MgII activity index
is shown in Figure\,1. The plotted $A_k$ values were determined on
the basis of the $a_{2k}$ coefficients calculated by Jesper Schou
for the SOHO MDI
data\footnote{http://quake.Stanford.EDU/~schou/asym1/}. The values
of $\Delta i_{\rm MgII}$ shown in the bottom panel are from M07
and they were based on the NOAA
data\footnote{http://www.sec.noaa.gov/ftpdir/sbuv/NOAMgII.dat}.

\begin{figure}
  \includegraphics[height=.48\textheight]{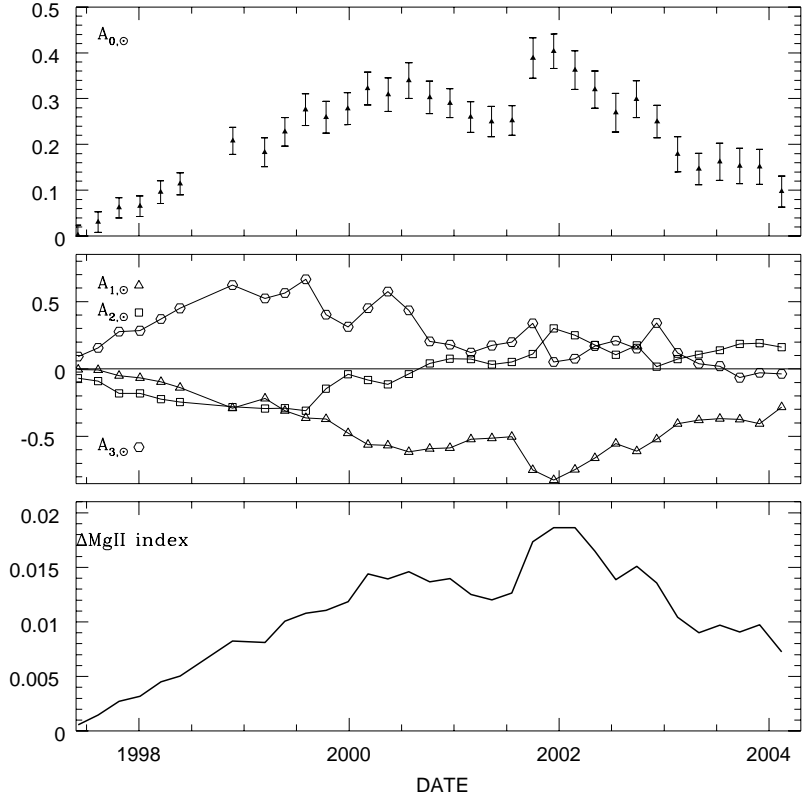}
  \caption{Low order $A_k$ coefficients describing the source of
  the oscillation frequency changes
  and the changes in the MgII activity index during activity cycle\,23.
  The coefficient $A_0$ with a $1\sigma$ error bar, shown in the
  top panel, describes the mean-frequencies shifts.
  The next three coefficients are shown in the middle panel.
  The errors bars are similar to those of $A_0$.}
\end{figure}

By comparing the top and lower panels in Figure\,1 (similar plots are
shown in Figure\,3 of M07), we see the excellent
correlation of the parameter $A_0$ with the changes in the MgII
index. Similar correlations exist for the CaII and other activity
indexes.
More complicated variability patterns of the higher order $A_k$, shown
in the mid
panel of Figure\,1, are closely related to the shift of the activity
from intermediate to low latitudes. We may see that there is a
correlation between $-A_1$ and $\Delta i_{\rm MgII}$ but other
coefficients reach their extreme values at different phases of the cycle.

Having determined the first four $A_k$-coefficients, we may calculate
the frequency shifts for modes with degrees $\ell\leq3$. Ignoring
small changes in the $a_i$ coefficients, related to the torsional
oscillations, we get from equations\,\eqref{Dn2} and \eqref{S2}
\begin{equation}
\Delta\nu_{nlm}
={R\over M}\sum_{k=0}^\ell A_kQ_{n\ell}(D_{c,k})\kappa_{k,lm},\label{Dn3}
\end{equation}
where $R$ and $M$ are expressed in solar units. We keep these
quantities, because we want to apply this equation to other stars.

With known $A_k(t)$, equation\,\eqref{Dn3} allows us to evaluate changes in
the mode frequencies within individual multiplets of low degrees. The plots in
Figure\,2 show the calculated frequency changes during cycle 23 of all
components in the two selected multiplets. The plots,
which begin at the solar minimum phase, take into account
rotational frequency shifts, including the linear shift of
$m\times0.45\mu$Hz and the tiny second-order effect arising from
the centrifugal distortion. At the minimum phase, the multiplets are
nearly symmetric about the $m=0$ mode frequency. The small
asymmetry of the $\ell=2$ multiplet comes mainly from the $a_4$
coefficient, which arises from the polar field. Even the smaller
asymmetry of the $\ell=1$ multiplet comes solely from the
centrifugal distortion. At the solar maximum the asymmetry is very
large. However, this spectacular signature of the magnetic
activity is not easy to detect. In fact the asymmetry of an $\ell=2$
multiplet was directly measured by Chaplin et al. (2003), long
after the first seismic signature of the solar activity has
been detected by Woodard and Noyes (1985).

\begin{figure}
  \includegraphics[height=.38\textheight]{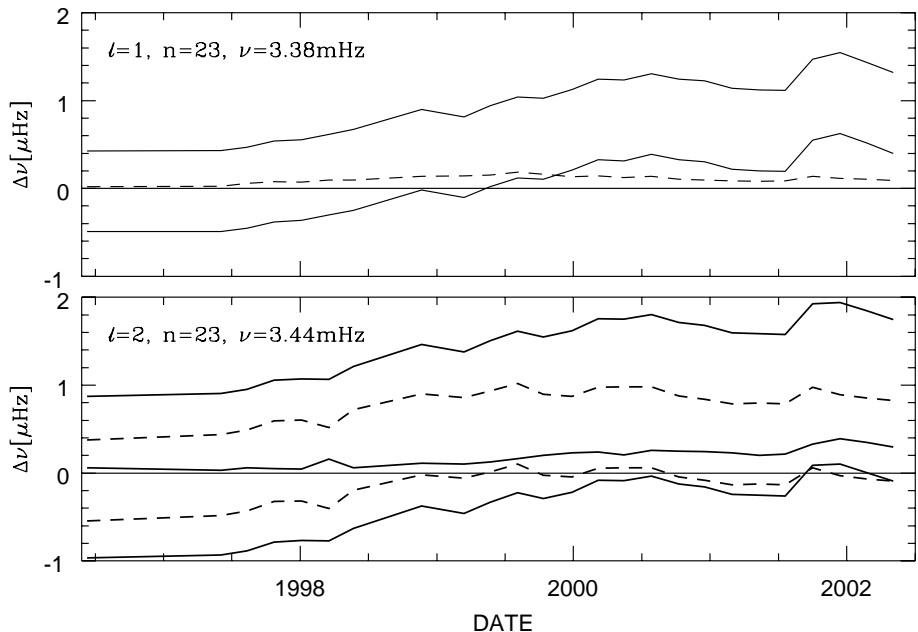}
  \caption{The changes in the multiplet structure during the activity
  cycle 23 for two
  selected multiplets calculated from the $a_0$, $a_1$, and $a_2$
  SOHO MDI data.
  The dashed lines refer to modes, which correspond to odd values of
  $\ell+|m|$ and thus are not detectable in the Sun. There is a drastic
  departure from
  symmetry about the zonal ($m=0$) frequency at high activity.}
\end{figure}

\section{Predictions for other stars}

The basic assumption in our prediction presented in M07 of
the $\ell=0$ mode frequency changes in $\beta$ Hyi was that the
relation $$A_0=22\Delta i_{\rm MgII},$$
which was found for various phases of the solar cycle is valid for other stars.
Instead of $i_{\rm MgII}$, one may use any other global activity index
available for
the Sun and for a considered star. The richest data on global activity
are from {\it the Mount Wilson Observatory - HK Project}, which
uses CaII. Unfortunately, this is a northern hemisphere project, which
neither includes $\beta$ Hyi, nor any other well studied solar-like
pulsators. To apply relation \eqref{Dn3}, we still had to scale
$D_{c,0}$ and for that we assumed that it is proportional to the
pressure scaleheight at the photosphere, that is
\begin{equation}
D_{c,0}\propto H_p\propto L^{0.25}R^{1.5}M^{-1}\label{Dc}.
\end{equation}

In order to make prediction for nonradial modes, we have to adopt
a scaling of the parameters at $k>0$ in equation\,\eqref{S2}. The
simplest choice is
\begin{equation}
A_k\propto A_0\propto\Delta i_{\rm MgII}\quad\mbox{ and }D_{c,k}\propto D_{c,0}.\label{AD}
\end{equation}
The first proportionality amounts to assume the same
{\it butterfly diagram} pattern as
in the Sun. The second is another extrapolation of the property found
in the solar data.
Once we make these additional assumptions, we may calculate, with the use
of equation\,\eqref{Dn3}, the expected frequency shift of low degree modes
for any star for which the shift in $i_{\rm MgII}$, or any other global
activity measure, is known.

If we had measurements of the individual mode frequencies within
multiplets, we could test our prediction based on equation\,\eqref{Dn3}
with single-season observations because the same equation
describes the momentary $\nu(m)$-dependence induced by the
momentary magnetic activity. However, such measurements may be
difficult, as we learnt from the case of the low-$\ell$ solar
oscillation. The problem is that the width of the peaks in the
oscillation spectrum is larger than the frequency splitting between
the modes. We thus expect that the mean (averaged over multiplet
components) frequency shifts will be the first detected seismic
signature of the magnetic activity in distant stars.

In order to calculate the mean frequency shifts, we have to
weight the individual components which depend
on the inclination of the rotation axis to the line of sight, $\theta_o$.
The weights should be proportional to the power, which --on average--
is proportional to $|Y_\ell^m(\theta_o,0)|^2$. Hence
(see equation\,\ref{Dn3}), the mean shifts are given by
\begin{equation}
\Delta\nu_{n\ell}={2\ell+1\over2}{R\over M}\sum_{k=0}^\ell
\left[\sum_{m=-\ell}^{m=\ell}|Y_\ell^m(\theta_o,0)|^2\kappa_{k,lm}\right]
A_kQ_{n\ell}(D_{c,k}).\label{Dn4}
\end{equation}
An unknown inclination introduces additional uncertainty to our
prediction for nonradial modes.

\section{The case of $\beta$ Hydri}

The G2IV star $\beta$ Hyi was chosen by M07 for predicting
radial mode frequency changes. Modes with $\ell\le2$
were measured in the two well separated seasons: 2000.5
(Bedding et al. 2001) and 2005.7 (Bedding et al. 2007). There are
eight identified p modes with $\ell\le2$ in both data sets.
Bedding et al. (2007) find only an insignificant mean frequency
shift of $0.1\pm0.4\mu$Hz between these dates.

Archival IUE spectra taken between 1978.5 and 1995.5 are available
for this star allowing to determine the changes in the MgII index. The
new analysis presented in M07 has led to an estimated cycle period of
12\,y and to a full amplitude of $\Delta i_{\rm MgII}=0.15$. Both
numbers are not very different from their solar counterparts. The
last observed activity minimum took place around 1992.5. In 2000.5
the star was about four years before the next estimated minimum and in
2005.7 about a year after. This means that the interval is not
well suited for testing our predictions but there is no better
data available on frequencies and activity for any object.

In order to apply equation\,\eqref{Dn4} for predicting frequency shifts, we
need a stellar model. The one used by M07 and
here has mass, $M=1.1$, radius, $R=1.82$, luminosity, $L=3.17$
(all in solar units), age=6.7 Gy, and the metallicity parameter
$Z=0.0138$. It reproduces the observed mean parameters of the star and
the large frequency separation. The values of $A_k$ and $D_c$ were
calculated according to relations \eqref{Dc} and \eqref{AD}, adopting
$\Delta i_{\rm MgII}=0.8\Delta i_{\rm MgII,\odot}$ and
$D_{c,\odot}=0.3\,$Mm. The numbers calculated for modes
identified in the $\beta$ Hyi oscillation spectrum by Bedding et al. (2007)
are shown in Figure\,3.
\begin{figure}
  \includegraphics[height=.4\textheight]{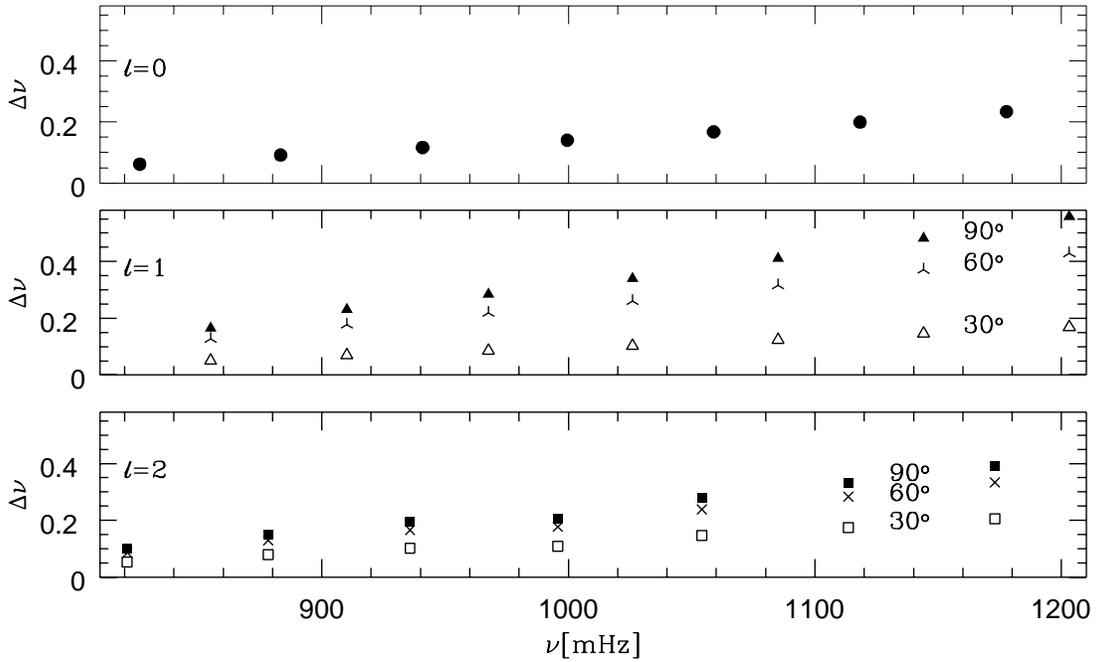}
  \caption{Predicted oscillation frequency shifts in $\beta$ Hyi between
   its activity minimum (2004.5) and maximum (2010.5). The dates were
   inferred in M07 on the basis of the $i_{\rm MgII}$
   behaviour between 1978.5 and 1995.5}
\end{figure}
From the numbers in this figure, we may see that within a few years we
have a chance to detect a marginally significant seismic signature
of the high activity state of $\beta$ Hyi. However, a major
improvement in frequency measurements is needed in order to learn
more about the star's activity changes by means of asteroseismology.
Once it is achieved and the differences in frequency shifts
between modes of different degrees are measured, we may get a
useful constraint on the longitudinal dependence of the activity
changes.

\section{Discussion}

The presented method of calculating an expected seismic signature of
stellar magnetic activity consists in extrapolating
the solar relation between changes in the p-mode frequencies and the
chromospheric activity indices. The extrapolation lacks a sound
physical justification so it should be regarded as a very
uncertain forecast.

Numerical predictions of the frequency shifts
were given for $\beta$ Hyi, which, like our Sun. is not a very
active star. The star is brighter and more evolved than the Sun but
there are no large differences between the two objects. The predicted shifts
were found similar to those measured in the Sun. They are small
and consistent with the null results of existing measurements but
they may be measurable within the next few years.

It could be more interesting to test the prediction for
a more active object. Unfortunately, there are none among those stars
for which data on solar-like oscillations are available. Magnetic activity
inhibits precise frequency measurements. Indeed, the analysis of the data
collected
during the solar cycle 23 shows that the widths of the peaks in the
oscillation spectra increase by some 15\%
at the high activity phase (Jim\'enez-Reyes et al. 2003).
On the other hand, since more active stars are
also faster rotators, the multiplet structure should be easier to
resolve. Thus, I believe such objects should be included in the
list of asteroseismic targets. The forecast based on the extrapolation
of solar data to more active stars is more risky.
However, regardless whether the forecast is right or wrong, its comparison
with reality will teach us something interesting about stellar activity.

\begin{theacknowledgments}
This work was supported by the Polish MNiI grant No.~1~P03D~021~28.
\end{theacknowledgments}

\end{document}